\documentclass[twocolumn,prl,showpacs]{revtex4}

\usepackage{graphicx}

\def\r{\vec r}
\def\u{\vec u}
\def\v{\vec v}
\def\a{\alpha}
\def\b{\beta}
\def\dab{\delta_{\alpha\beta}}
\def\dt{\delta t}
\def\dr{\delta r}

\begin{document}

\title{Lattice Boltzmann Method with regularized non-equilibrium distribution functions}

\author{Jonas Latt and Bastien Chopard}
\affiliation{%
Computer Science Department, University of Geneva\\
CH-1211 Geneva 4, Switzerland
}%

\date{April 24, 2005}

\begin{abstract}
A new lattice Boltzmann (LB) model is introduced, based on a regularization of the pre-collision distribution functions in terms of the local density, velocity, and momentum flux tensor. The model dramatically improves the precision and numerical stability for the simulation of fluid flows by LB methods. This claim is supported by simulation results of some 2D and 3D flows.
\end{abstract}

\pacs{47.11.+j, 05.20.Dd}

\maketitle

The lattice Boltzmann (LB) model is a recent technique for the simulation and modeling of fluid flows~\cite{succi-book:01,wolf-gladrow:00,BC-acs:02,chen-doolen:91,review:03}. During the past fifteen years it has been successfully applied to many challenging problems in hydrodynamics as well as reaction-diffusion processes and wave propagation phenomena~\cite{BC-livre}. A particular shortcoming of this technique are numerical instabilities, that may develop at high Reynolds numbers. Several improvements to the method have been proposed, which however either induce a substantial complication of the original algorithm, or require a cumbersome fine-tuning of adjustable parameters~\cite{lallemand:00,ansumali:02}. The present paper introduces a new method which fits quite naturally into the framework of classical LB models and offers both increased accuracy and stability at very low cost.

The LB approach considers a mesoscopic description of the fluid on a regular lattice of spacing $\dr$ in $d$-dimensions. The central quantities of the LB approach are distribution functions $f_i(\r,t)$, which denote the density of particles entering a lattice site $\r$ at discrete time $t$ with velocity $\v_i$.  The $\v_i$ are vectors connecting any lattice site $\r$ with its $z$ neighbors $\r+\dt\v_i$, $\dt$ being the time step and $z$ the lattice coordination number. A vector $\v_0=0$ corresponding to a rest population $f_0$ is also introduced. The LB dynamics are expressed as
\begin{equation}
f_i(\r+\dt\v_i, t+\dt)=f_i(\r,t) + \Omega_i(f(\r,t)),
\label{eq:LB-general}
\end{equation}
where $i$, here and in subsequent formulas, runs from 0 to $z$. The dynamics can be split conceptually into a collision step by defining $f_i^{out}=f_i+\Omega_i (f)$, and a propagation step: $f_i(\r+\dt\v_i, t+\dt)=f_i^{out}(\r,t)$. During the collision step, the advected particle streams $f_i$ are summed up with the collision terms $\Omega_i$, which are given functions of the $f_i$'s. They describe how fluid particles colliding at site $\r$ change their velocities to $v_i$. Then, at the propagation step, the fluid particles are streamed to the neighboring site $\r+\dt\,\v_i$.

As in any standard kinetic theory, the macroscopic quantities are
obtained by taking the first velocity moments of the distribution functions:
\begin{equation}
\rho=\sum_{k=0}^z f_k, \ \rho\u=\sum_{k=0}^z f_k\v_k, \ \Pi_{\a\b}=\sum_{k=0}^z f_k v_{k\a}v_{k\b},
\label{eq:macro-qtty}
\end{equation}
where $\rho$, $\u$, and $\Pi$ are the fluid density, momentum, and
momentum flux tensor respectively (Note that the actual momentum flux tensor in
LB models has an extra lattice contribution, which adds on to $\Pi$.).  Here and in what follows, Greek indices
label the components of two-dimensional (2D) resp. three-dimensional (3D) physical space, whereas Latin indices refer to the $z+1$-dimensional space of the distribution functions. Vectors situated in the former space are characterized by an arrow on top of the letter, and in the latter space, simply by omitting the index.

The collision term $\Omega$ is chosen in such a manner that mass and momentum are conserved exactly (without discretization error), so as to closely reflect the physical laws at the base of hydrodynamics.
Its most common implementation, the BGK model, expresses a single-time relaxation to a given local
equilibrium function $f^{eq}$, depending only on the conserved
quantities $\rho$ and $\u$ calculated from~(\ref{eq:macro-qtty}):
\begin{equation}
\Omega_i=-\omega \left(f_i - f_i^{eq}(\rho,\u)\right),
\end{equation}
where $0<\omega <2$ is the relaxation parameter, directly related to the dynamic fluid viscosity $\nu$.

The expression for $f^{eq}$ comes from a low Mach number truncated Maxwell-Boltzmann distribution and is adjusted to obtain the correct momentum flux tensor:
\begin{equation}
\Pi_{\a\b}^{eq}=\sum_{k=0}^z f_k^{eq}v_{k\a} v_{i\b}=
   \rho c_s^2\dab + \rho u_\a u_\b,
\label{eq:Pi-eq}
\end{equation}
where $c_s$ is the speed of sound.
The equilibrium term $f^{eq}$ reads~\cite{succi-book:01,wolf-gladrow:00,BC-acs:02}
\begin{equation}
f_i^{eq}=\rho t_i\left[ 1 + {v_{i\a}u_{\a} \over c_s^2} 
                  + {1\over 2c_s^4}Q_{i\a\b}u_\a u_\b \right],
\label{eq:feq}
\end{equation}
where a repeated Greek index implies a summation over this index. The tensors
$Q_{i\a\b}$ are defined to be $Q_{i\a\b}=v_{i\a} v_{i\b} -c_s^2\dab$,
and the $t_i$'s, as well as $c_s$, are coefficients specific to the lattice topology.

The connection between the LB method and the corresponding hydrodynamics is obtained
through a Taylor expansion, up to second order in $\dt$, of the finite
differences in the left hand side of Eq.~(\ref{eq:LB-general}),
and a multiscale Chapman-Enskog expansion $f = \sum_{k=0}^{\infty} f^{(k)}$. The zeroth-order term yields the equilibrium distribution value $f^{(0)}=f^{eq}$, and the remaining terms are denoted as $f^{neq}$:
\begin{equation}
f_i^{neq} = f_i - f_i^{eq} \quad \textrm{and} \quad \Pi^{neq} = \Pi-\Pi^{eq}.
\label{eq:fneq}
\end{equation}
For the BGK model, the first-order multiscale Chapman-Enskog procedure
gives~\cite{BC-acs:02}
\begin{equation}
f_i^{neq}\approx f_i^{(1)}=-{\dt \over \omega c_s^2}t_i
Q_{i\a\b}\partial_\a \rho u_\b,
\label{eq:f1}
\end{equation}
and we obtain
\begin{equation}\label{eq:Pi-neq}
\Pi_{\a\b}^{neq} \approx \sum_{k=0}^z f_k^{(1)}v_{k\a} v_{k\b} = -{\dt c_s^2 \over \omega} \left(\partial_\alpha \rho u_\beta + \partial_\beta \rho u_\alpha \right).
\end{equation}
Using expressions~(\ref{eq:Pi-eq}) and (\ref{eq:Pi-neq}) together with
the lattice contribution to the momentum flux (see for
instance~\cite{BC-acs:02}), it can be shown that $\u$ obeys the
Navier-Stokes equation with the viscosity given by
\begin{equation}
 \nu=\dt\, c_s^2\left(\frac{1}{\omega}- \frac{1}{2}\right).
\label{eq:viscosity}
\end{equation}

However, in actual numerical simulations, the proposed theoretical
description of the LB dynamics is not fully obeyed because $\dt$ and
$\dr$ are not arbitrarily small, and also because higher order
derivatives are neglected in the approximation~(\ref{eq:f1}). As a
result, the numerical behavior departs from its hydrodynamic limits
and numerical instabilities may appear if some quantities vary too
sharply over time and space.  

\begin{figure}
\includegraphics[width=\columnwidth]{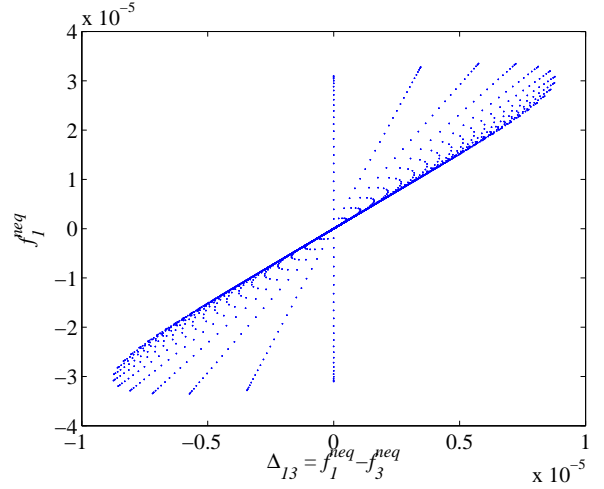}
\caption{\label{fig:f1-versus-f3}Difference between non-equilibrium parts of distribution functions along two opposite directions $\v_1/v=(1,0)$ and $\v_3/v=(-1,0)$. Each data point is obtained from one lattice site on a numerical simulation of a Kovasznay flow, at $Re=1$.}
\end{figure}

The inaccuracy of the first-order terms $f^{(1)}$ becomes apparent, {\it e.g.}, upon the observation that, according to Eq.~(\ref{eq:f1}), $f^{(1)}$ is symmetric with respect to spatial reflections: the difference $f_i^{(1)} - f_j^{(1)}$ vanishes along directions $i,j$ for which $v_i = -v_j$. In practice, this relation is not necessarily obeyed by the non-equilibrium parts of the distribution functions. On Fig.~\ref{fig:f1-versus-f3}, $\Delta_{ij} = f_i^{neq}-f_j^{neq}$ is plotted, for a given couple $\{i,j\}$, on ground of some numerical data of the Kovasznay flow described below. It appears to take nonnegligible values at the scale of non-equilibrium terms (up to $30\%$).

To reduce the discrepancy between $f^{neq}$ and $f^{(1)}$, we propose
a regularization procedure whose goal is to force the numerical scheme
to comply as much as possible with the theoretical framework exposed
above. For this purpose we recompute $f^{neq}$ prior to the
collision step so as to enforce $f^{neq}=f^{(1)}$. The key of our
regularization procedure is the observation that Eqs.~(\ref{eq:f1})
and~(\ref{eq:Pi-neq}) can be combined to give
\begin{equation}
f_i^{(1)}={t_i\over 2c_s^4}Q_{i\a\b}\Pi_{\a\b}^{neq}.
\label{eq:f1-versus-Pi}
\end{equation}

In conclusion, our regularization procedure amounts to computing
the regularized values $f^{(1)}$ of $f^{neq}$ according to the
following steps:
\begin{eqnarray}
f &\stackrel{Eq.~(\ref{eq:macro-qtty})}{\longrightarrow}& \left( \begin{array}{c} \rho\\ \rho\u \end{array}\right)
              \stackrel{Eq.~(\ref{eq:feq})}{\longrightarrow} f^{eq}(\rho,\u) \label{eq:reg-steps} \\
  &\stackrel{Eq.~(\ref{eq:fneq})}{\longrightarrow}&  \left( \begin{array}{c} f^{neq}\\ \Pi^{neq} \end{array}\right)  \stackrel{Eq.~(\ref{eq:f1-versus-Pi})}{\longrightarrow} f^{(1)} \nonumber
\end{eqnarray}
Then, the standard BGK collision is applied to $\bar{f}=f^{eq}+f^{(1)}$, and the
regularized collision step of the dynamics reads
\begin{equation}
f_i^{out}=f_i^{eq} + (1-\omega) f_i^{(1)}.
\label{eq:BGK-regularized}
\end{equation}
Note that since $\sum_k Q_{k\a\b}=\sum_k Q_{k\a\b}\v_k=0$, the above
scheme still conserves mass and momentum exactly . 

In order to better understand the way the steps described in Eq.~(\ref{eq:reg-steps}) act on the distribution functions, it is illuminating to study the dynamics in the ($z+1$)-dimensional space of the velocity moments. These moments [a few of them are shown in Eq.~(\ref{eq:macro-qtty})] are associated in kinetic theory with so-called modes of the collision operator and can be related to transport phenomena during the collision process. In general, the moment space is related to the space of the distribution functions through an invertible linear mapping whose matrix $M$ is explicited, {\it e.g.}, in~\cite{lallemand:00}. The regularized dynamics presented in Eqs.~(\ref{eq:reg-steps}, \ref{eq:BGK-regularized}) can be reformulated as
\begin{equation}\label{eq:reg-moment}
 f^{out} = f^{eq} + (1-\omega)\, M^{-1} A M f^{neq},
\end{equation}
where $A = M\,R\,M^{-1}$, with $R_{ij} = \frac{t_i}{2 c_s^4} Q_{i\a\b}c_{j\a}c_{j\b}$.
In 2D and under the assumption of fluid incompressibility $Tr(\Pi)=0$, the matrix $A$ is found to be diagonal: $A=diag(\lambda_0,\lambda_1,\cdots,\lambda_z)$, where $\lambda_{i_1}=\lambda_{i_2}=1$ for the components of the momentum flux tensor\footnote{Only two of the three independent components of the 2D momentum flux tensor are represented in the base of moment space, the (compressible) component $Tr(\Pi)$ being linearly dependent on other moments.} and $\lambda_i=0$ for the other moments. In the general (compressible) case, additional off-diagonal contributions appear in the energy and square-energy moments. This interpretation of the dynamics shows that, except for compressibility effects, the regularized dynamics directly kills all modes but the ones associated to the momentum flux tensor.

It is interesting to compare the regularized method with so-called multi-relaxation-time (MRT) models, \cite{dhumieres:92, lallemand:00}, which propose the following general formulation of the LB dynamics:
 $
 f^{out} = f(\r,t) - M^{-1}S(\mathcal{F}-\mathcal{F}^{eq}),
 $
where $\mathcal{F}=M f$ is the moment space representation of the distribution functions, $S$ is a diagonal matrix $S=diag(s_0,s_1,\cdots,s_z)$ containing $z+1$ individual relaxation parameters $s_i$, and $\mathcal{F}^{eq}$, the equilibrium distribution in moment space, depends on a set of adjustable parameters. By fixing those adjustable parameters through the relation $\mathcal{F}^{eq}=M f^{eq}$ and the relaxation parameters through $s_i=\omega$ for all non-conservative momenta, the usual BGK dynamics are recovered. It has however been argued~\cite{lallemand:00}, that the stability of the BGK scheme is enhanced by an appropriate choice of the various relaxation parameters.

In the case $\mathcal{F}^{eq}=M f^{eq}$, the MRT model takes the following form:
$
 f^{out} = f(\r,t) - M^{-1}SM(f-f^{eq}),
 $
which, in analogy with Eq.~(\ref{eq:reg-moment}), can be reformulated as
\begin{equation}\label{eq:mrt-reformule}
 f^{out} = f^{eq} + ({\bf 1} - M^{-1} S M) f^{neq}.
\end{equation}
Here, the identity term $\bf 1$ is due to the advected distribution functions, which are not touched upon by the MRT correction to the BGK model. Eqs.~(\ref{eq:reg-moment},~\ref{eq:mrt-reformule}) make the main difference between our regularized model and the MRT approach apparent: while in the MRT approach [Eq.~(\ref{eq:mrt-reformule})] non-physical modes are relaxed to a local equilibrium inside the collision term, in the regularized model [Eq.~(\ref{eq:reg-moment})] these modes are more radically eliminated in both the advected particles and the collision term. Therefore, when increasing the stability in the simulation of a Navier-Stokes fluid flow is the only issue, our method is comparatively simpler from a theoretical viewpoint, and efficient to implement.

\begin{figure}
\includegraphics[width=\columnwidth]{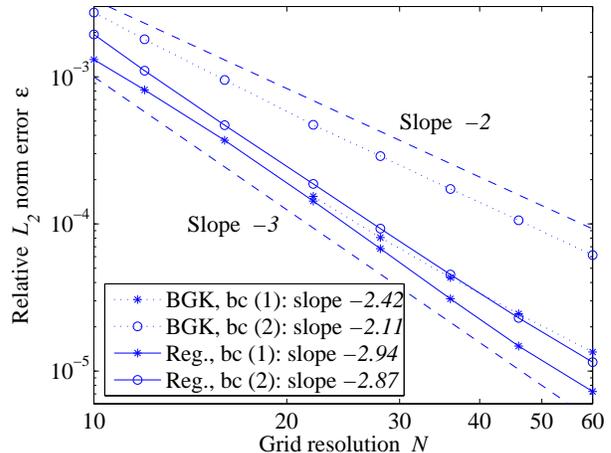}
\caption{\label{fig:wake-kovasznay}Relative error of numerical result on Kovasznay flow in wake region. Both traditional BGK and the regularized model are tested on two common boundary conditions.}
\end{figure}
We now turn to numerical verifications of the regularized model on two 2D flows using a D2Q9 lattice, and one 3D flow using a D3Q19 lattice~\cite{succi-book:01,wolf-gladrow:00,BC-acs:02}. The first test concerns the simulation of a Kovasznay flow, which approximates the stationary 2D flow behind a regular grid. An analytical solution for this flow, proposed in~\cite{kovasznay:48}, takes the following form:
\begin{eqnarray}
u_x &=& u_\infty(1-\exp(\lambda x/L)\cdot \cos(2\pi y/L))\quad\textrm{and}\label{eq:kovasznay}\\
u_y &=& u_\infty\frac{\lambda}{2\pi}\exp(\lambda x/L)\cdot \sin(2\pi y/L),\quad \textrm{with}\nonumber\\
\lambda &=& Re/2 - \sqrt{4\pi^2+Re^2/4},\nonumber
\end{eqnarray}
where $u_\infty$ is the asymptotic velocity of the fluid, $Re = u_\infty L/\nu$ is the Reynolds number, and $L$ defines the length scale of the problem. The simulations are performed in the wake of the grid, in the intervals $x \in [L/2, 2L]$ and $y \in [-L/2, 3L/2]$, with $Re=10$, $u_\infty=0.01\,v$, and with a varying grid resolution $N=L/\dr$. Keeping the velocity constant in terms of the lattice unit $v$ amounts to fixing the Mach number $Ma=u/c_s$ at a value sufficiently small to mimic an incompressible flow. Given that the flow is periodic in $y$-direction, the upper and the lower boundary of the simulation can be chosen periodic, whereas the Kovasznay solution [Eq.~(\ref{eq:kovasznay})] is imposed through Dirichlet boundary conditions on the left and right boundary. After the simulation has stabilized, the numerical result is compared with the solution [Eq.~(\ref{eq:kovasznay})] through an $L_2$ norm on each grid point, and then averaged over space. The result is shown in Fig.~\ref{fig:wake-kovasznay}, on two commonly used implementations of the boundary conditions (bc); bc~(1) \cite{inamuro:95} and bc~(2) \cite{skordos:93}. The accuracy of the simulation with respect to the grid resolution is of order $2$ to $2.5$ when the BGK model is used, whereas the regularized model is almost third-order accurate. On the BGK simulations with bc~(1), data points for small grids are missing because numerical instabilities make them impossible, whereas the regularized model has no such stability deficiencies.

\begin{figure}
\includegraphics[width=\columnwidth]{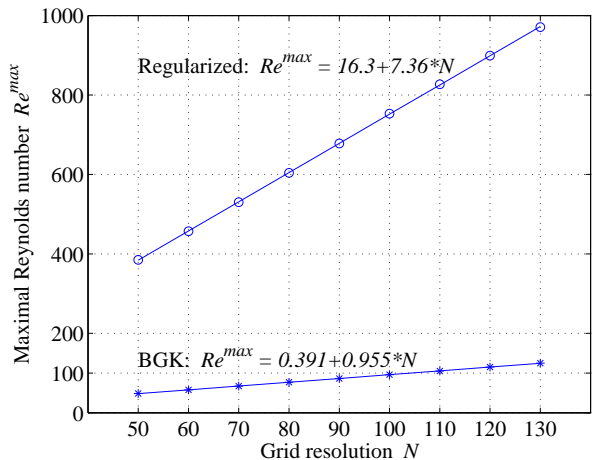}
\caption{\label{fig:cavity}Simulation of 2D cavity flow for fixed Mach number. $\circ$, $\ast$: maximal stable Reynolds number, numerically determined; solid line: least-square linear fit of the data points (parameters of the fit are indicated on the graph).}
\end{figure}
The second test case implements a flow in a 2D square cavity whose top-wall moves with a uniform velocity. Both standard BGK and the regularized model are first compared with the reference solution of Ghia {\it e.a.}~\cite{ghia:82}, on a lattice size of $N\times N$ with $N=129$, at $Re=100$ and a top-wall velocity $u_0=0.02\,v$. A boundary condition described in~\cite{zou_he:97} is used. The reference solution~\cite{ghia:82} proposes a set of accurate numerical values for some $x$- and some $y$-components of the velocity on chosen space points. An $L_1$ norm error with respect to these reference points is averaged over all available points and normalized with respect to $u_0$. For the BGK model, this yields an error of $\epsilon=3.71\cdot10^{-3}$, and for the regularized method, of $\epsilon=2.40\cdot10^{-3}$. Thus, both methods solve the problem with satisfying accuracy. The regularized model is however found to be substantially more stable. To make this statement more quantitative, a series of simulations is run, on which the velocity (and thus the Mach number) is kept constant at $u_0=0.02 v$. For several chosen grid sizes $N$, the maximal Reynolds number $Re^{max}$ at which the simulation remains stable ({\it i.e.} delivers finite numerical values) is determined. Figure~\ref{fig:cavity} shows that, although both methods exhibit a linear relationship between $Re^{max}$ and $N$, the observed increase rate is $7.7$ times higher for the regularized method than for BGK.

Finally, the capacity of the regularized model to represent 3D flows has been explored in a preliminary study on direct numerical simulations (DNS) of a homogeneous and isotropic turbulent flow. The system possesses periodic boundaries and is driven by an external force that excites two wavenumbers in the limit of large wavelengths~\cite{alvelius_a:99, kate:02}. It is known that the energy injected in such a system is mainly dissipated at the smallest scales, whose size is estimated by the so-called Kolmogorov length $l_k$. If these scales are not resolved with sufficient accuracy in the simulation, the system accumulates the energy and develops numerical instabilities. Our numerical simulations show that indeed, when the Kolmogorov length is of the order of magnitude of a lattice site, $l_k=0.5\, \dr$, with an average velocity $\bar u = 0.04\, v$, both BGK and the renormalized model exhibit a numerically stable flow. Furthermore, their statistical properties are numerically verified to fit the predictions of the theory of fluid turbulence. However, at a smaller Kolmogorov length (and thus higher $Re$) $l_k=0.06\,\dr$, BGK is numerically unstable, whereas numerical stability is still ensured by the renormalized model. This observation suggests that the physics of the small scales are represented more accurately by the renormalized model than by BGK.

In this paper, a novel numerical scheme has been presented for the simulation of fluid flows by the LB method. It has been compared with the traditional BGK method and shown to be substantially more precise on a problem with mathematically well defined boundaries, dramatically more stable on a problem with high pressure gradients on a critical point, and more robust against an excessive energy input in a turbulent flow. Given its conceptual simplicity, we highly recommend its use as an alternative model for the simulation of complex fluid flows. We thankfully acknowledge the support by the Swiss National Science Foundation~(SNF).

\end{document}